\documentclass{mem}
\usepackage{natbib}\usepackage{txfonts}\usepackage{balance}
\usepackage{graphicx}
\usepackage[a4paper]{hyperref}
\idline{1}{1}
\def\int{{\it INTEGRAL}}
\def\rxs{1RXS J170849.0--400910}
\def\zerosei{SGR 1806--20}
\begin{document}

\title{
Hard X-ray Emission from Magnetars
}

   \subtitle{A Case Study for Simbol X}

\author{
Diego G\"otz\inst{1} 
          }

  \offprints{Diego G\"otz}

\institute{
CEA Saclay, DSM/Dapnia/Service d'Astrophysique, F-91191, Gif sur Yvette, France
\email{diego.gotz@cea.fr}
}

\authorrunning{G\"otz}

\titlerunning{Hard X-ray Emission of Magnetars}

\abstract{
The magnetar model involves an isolated neutron star with a very high magnetic field ($B\sim 10^{14-15}$ G), and is invoked to explain the emission processes of two classes of sources, the Anomalous X-ray Pulsars (AXPs) and the Soft Gamma-Ray Repeaters (SGRs). Five of them have been recently identified to be persistent sources in the hard X-ray band (20--200 keV). AXPs, in particular, present the hardest known persistent spectra in the hard X/soft $\gamma$-ray energy range. The broad band modeling of their spectra still suffers from the non-simultaneity of the observations and from a lack of sensitivity above 20 keV. We present the Simbol X simulated observations of these objects and show that that this mission could surely help to disentangle the contribution of the different spectral components, and to understand how they contribute to the secular flux variations observed in these sources.
}
\maketitle{}

\section{Introduction}

Most of the known neutron stars (NSs) are either isolated rotation powered pulsars, or accretion powered neutron stars
hosted in binary systems. A dozen (plus a few candidates) of sources, dubbed magnetars, do not fit in either of these categories, since their dominant source of energy is believed to be the magnetic one. In fact, in the magnetar model 
\citep{dt92} it is the decay of the huge magnetic field ($B\sim 10^{14-15}$ G) of isolated neutron stars that powers their electromagnetic emission. 

Magnetars are historically divided in two categories, due to the fact that they were discovered in different ways. We will briefly summarize their properties here. For a complete review of this kind of objects see \citet{woodsrew}.

\subsection{Soft Gamma-Ray Repeaters}

SGRs were originally identified in the late '70s a subclass of Gamma Ray Bursts, due to the fact that these bursts had softer spectra and originated repeatingly from fixed positions across the sky. Four confirmed SGRs are known to date, three of them (1806--20, 1900+14, and 1627--41) are located in our Galaxy at several kpc, while one, 0525--66, is located in the Large Magellanic Cloud. They sporadically emit short ($\sim$0.1 s) and bright (L$\sim$ 10$^{39}$-10$^{42}$ ergs s$^{-1}$) bursts of soft gamma-ray radiation during active periods, alternating with quiescent periods that can last several years. In rare occasions they emit so-called {\em giant flares}, which are characterized by very brights spikes (L$\sim$ 10$^{44}$-10$^{46}$ ergs) lasting a fraction of a second, followed by a pulsating tail lasting a few hundred seconds. Up to know three out of four SGRs have emitted a giant flare, and the pulsations found in their tails pushed forward the idea that these objects were associated with neutron stars, with an extremely high magnetic field. In fact,
the energy required to confine a fireball for the duration of the giant flare pulsating tail strongly points
towards a magnetic field of the order of 10$^{15}$ G.

Three of the confirmed SGRs have quiescent pulsating X-ray counterparts for which the period and its derivative could be measured. Assuming that this radiation is due to magnetic braking in a dipolar surface magnetic field, these values confirm the high magnetic field, and the absence of Doppler modulations in the observed light curves exclude the presence of companion stars.

\subsection{Anomalous X-ray Pulsars}

AXPs have been originally identified as a class due to their common X-ray and timing properties \citep{merestella}.
Their rotational periods cluster in the 5-12 s range, their period derivatives in the 0.05-4$\times$10$^{-11}$ s s$^{-1}$.
Their X-ray luminosity is L$_{X}\sim$10$^{34-36}$ erg s$^{-1}$ . From their timing porterties one can derive the dipole magnetic field at the surface of the NS as follows, 

\begin{equation}
B=\left(\frac{3Ic^{3}P\dot P}{2\pi^{2}R^{6}}\right)\simeq3.2\times10^{19}(P\dot P)^{1/2} \; \rm G, 
\end{equation}

where $I$ ($\simeq$10$^{45}$g cm$^{2}$) is the NS moment of inertia, and $R$ ($\simeq$10$^{6}$ cm) is the NS radius. The derived field values for AXPs exceed the quantum critical value of $B_{Q}\equiv m_{e}^{2}c^{3}/(e\hbar)=4.4\times10^{13}$ G. This fact together with the lack of an observable companion star and with the fact that $L_{X}$ largely exceeds the rotational energy loss ($I\omega\dot\omega$), points towards the inclusion
of the AXPs in the magnetar class. In addition
the detection of Soft Gamma-Ray Repeater-like bursts from five AXPs
\citep[e.g.][]{gavrill} has strengthened the association between these objects and the SGRs.

\section{INTEGRAL Results}

\begin{figure}[ht!]
\resizebox{\hsize}{!}{\includegraphics[width=7cm]{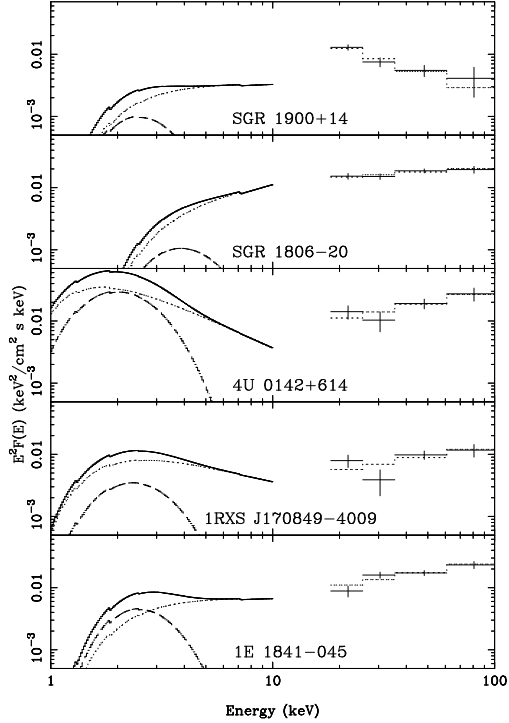}}
\caption{\footnotesize 
Broad band X--ray spectra of the five magnetars detected by \int~. The
data points above 18 keV are the \int~ spectra with their best fit
power-law models (dotted lines). The solid lines below 10 keV represent
the absorbed power-law (dotted lines) plus blackbody (dashed lines) models
taken from the literature. See \citet{dg06} and references therein for details.
}
\label{fig:int}
\end{figure}

The soft X-ray spectra of AXPs and SGRs are generally well described by a two
component model, made of a black body with $kT\sim$ 0.4--0.5\,keV,
and a steep power law, with photon index $2\leq\Gamma \leq 4$.
Being their spectra below $\sim$10 keV rather soft, the
first detections above 20 keV of very hard high-energy tails
associated with these objects came as a surprise
\citep{kuiper,denhartog,rev,mere05,molkov,dg06}. These discoveries were possible thanks to
unprecedented sensitivity of the IBIS/ISGRI imager \citep{ibis,isgri} on board the \int~ satellite \citep{integral}.
It results that AXP spectra flatten ($\Gamma\sim$ 1)
above 20 keV (and the pulsed fraction of some of them reaches up to 100\% \citep{kuiper}),
while SGR spectra steepen at hard X-rays, as is illustrated in Fig. \ref{fig:int}. 

The discovery of these hard tails provides new constraints on the emission models for these objects
since their luminosities might well be dominated by hard, rather than soft, X-rays.
In fact, quite recently, \cite{tb05} discussed how soft gamma-rays may be
produced in a twisted magnetosphere, proposing two different
scenarios: either thermal bremsstrahlung emission from the surface
region heated by returning currents, or synchrotron emission from
pairs created higher up ($\sim$ 100 km) in the magnetosphere.
While both scenarios predict a power-law-like spectral
distribution for the 20-100 keV photons, the cut-offs of the high
energy emission are markedly different in the two cases, 100 keV
vs. 1 MeV. A third scenario involving resonant magnetic
Compton up-scattering of soft X-ray photons by a non-thermal
population of highly relativistic electrons has been proposed by
\cite{baring}.

\begin{figure}[ht!]
\resizebox{\hsize}{!}{\includegraphics[angle=90,width=7cm]{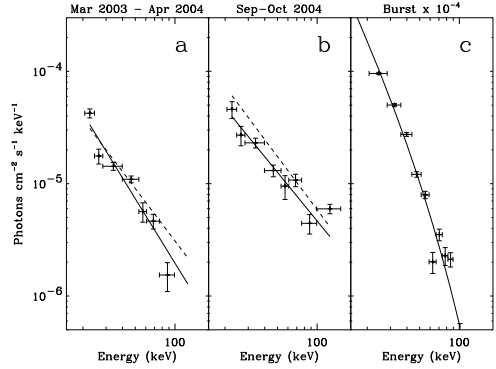}}
\caption{\footnotesize 
IBIS/ISGRI spectra of \zerosei. a) persistent emission March 2003-April 2004, b) persistent emission September-October 2004, c) one burst (scaled down by a factor 10$^{4}$). The solid lines are the best fits (power laws in a) and b), thermal bremsstrahlung in c)). The dashed lines indicate the extrapolation of power-law spectra measured in the 1--10 keV band with {\em XMM-Newton} \cite{xmm1806}. From \cite{mere05}.
}
\label{fig:1806}
\end{figure}

\begin{figure}[ht!]
\resizebox{\hsize}{!}{\includegraphics[angle=-90,width=7cm]{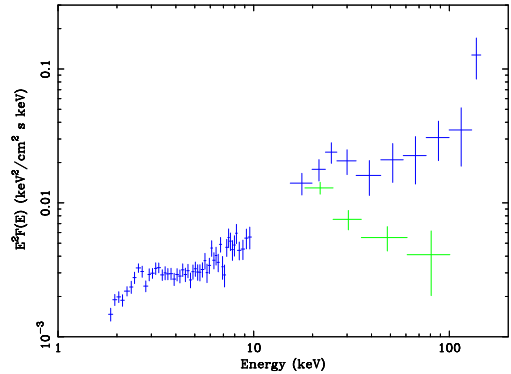}}
\caption{\footnotesize 
Blue points: broad band spectrum of \mbox{SGR\,1900+14} taken on 1997 May 12 (observation A) with {\it BeppoSAX} (both MECS and PDS data). Green points: \emph{INTEGRAL} data from March 2003 to June 2004. From \citet{esposito1900}
}
\label{fig:1900}
\end{figure}

While at soft X-rays AXPs and SGRs have shown to be variable objects, as far as timing and spectral properties are concerned, at hard X-rays the study of these characteristics is limited by the long integration times required in
order to detect them with the current instrumentation. Nevertheless, some degree of variability in the hard X-ray emission of SGRs has been reported recently by different authors.
\citet{mere05} report, besides clear flux variations, a possible spectral hardening of \zerosei~ as a 
function of its bursting activity (see Fig. \ref{fig:1806}). Similar results have been obtained by \citet{esposito1900}, who compared the broad band persistent spectrum of SGR 1900+14
measured with {\it Beppo}SAX in 1997 during a period of intense bursting activity with the one taken with
IBIS/ISGRI six years later, during a phase of quiescence (see Fig. \ref{fig:1900}). The latter represents for the time being
the most convincing evidence of spectral variability in an SGR hard tail.
\section{Simbol-X Simulations}
As mentioned above, the limited sensitivity of the current instrumentation does not allow us to study simultaneously
the broad band spectra of magnetars. In fact, while at soft X-rays a 50 ks observation using focusing telescopes
like {\it XMM} and {\it Chandra} are sufficient to derive precise spectral and timing characteristics, above 20 keV 
IBIS/ISGRI, the most sensitive instrument in this energy range nowadays, requires at least 500 ks to derive a useful
spectrum, often still with large uncertainties. In addition due to the mission constraints the observation has to be
split over several months. In these conditions it is clear that it is difficult to speak of {\em simultaneous} broad band
spectra. This is a severe limitation to any sensible spectral modeling: for instance the three components identified
in the AXPs spectra, the black body, the low-energy power law and the high energy one seem for the time being completely independent spectral components. In particular very little known concerning the transition region between
10 and 20 keV where no data is actually available. This gap, together with the unavoidable uncertainties in the
inter-calibration of the different instruments increases our difficulties in deriving a physical model of the source
that accounts simultaneously for the three components, and explains what is the physical driver of the measured
spectral and flux variations.

Stimulated by these issues I tried to figure out if Simbol-X \citep{ferrando} 
could contribute to disentangle some of these aspects.
This was done through some simulated observations performed with the {\it xspec} package, and following the
recipes provided for this workshop by Jean Luc Sauvageot. The simulations included the low energy layer MPD (Marco Pixel Detector) and the high energy CZT, in a 20 m focal length configuration and 12 arcmin field of view with a 18 arcsec point spread function at 30 keV. For the background estimation the contributions of the diffuse X-ray background, the local 
bubble, and the cosmic rays interactions (the detectors are actively and passively shielded) have been taken into account.
 
\subsection{\rxs}

\begin{figure}[ht!]
\resizebox{\hsize}{!}{\includegraphics[angle=-90,width=7cm]{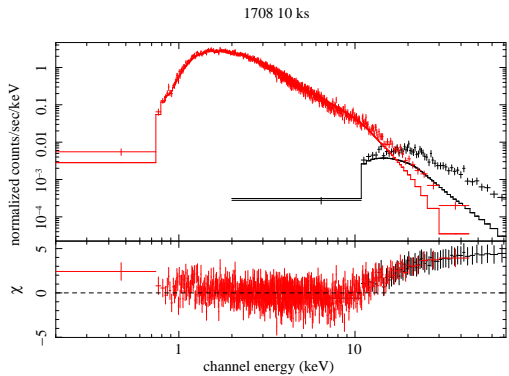}}
\caption{\footnotesize 
10 ks simulated Simbol X observation of \rxs. The upper panel shows the model and the data while the lower panel shows
the residuals with respect to the model. The data have been fitted with the model derived from soft X-ray data only.}
\label{fig:1708_10k}
\end{figure}

\rxs~ is one of the persistent AXPs that shows the largest degree of variability in its spectral parameters \citep[see e.g][]{campana}. I wanted to estimate the observation time needed in order to carefully characterize the spectral components of this
AXP with Simbol-X. In order to do this I simulated a spectrum made of the sum of an absorbed black body and a broken power-law changing its slope from 2.8 to 1.5 at 15 keV. Such a sharp break may not be physical but is consistent
with the available spectral data. From Fig. \ref{fig:1708_10k} one can see that in just 10 ks of observation time
one can get a fairly good spectrum. In the figure I show the fit using an absorbed black body and a single power
law. The fit is unacceptable ($\chi_{r}^{2}\sim 2$), and the break is clearly detected in the residuals at 15 keV.

This shows that even with a relatively short observation we can derive a good broad band spectrum with Simbol-X, and
that we will be able to monitor the spectral changes of these sources on a time scale that is not accessible with the current instrumentation.

\subsection{\zerosei}

%
%
%
%

For SGRs, and for \zerosei~ in particular, the presence of a spectral break around 15 keV is not settled yet. The available
data suffer from non-simultaneity and inter-calibration uncertainties. As a working hypothesis I assumed a spectrum
made again of an absorbed black body plus a broken power law, but this time the black body component is much
fainter, as observed in this source by {\it XMM} \citep{xmm1806}, and the slope changes from 1.2 to 1.8 at 15 keV.

\begin{figure}[ht!]
\resizebox{\hsize}{!}{\includegraphics[angle=-90,width=7cm]{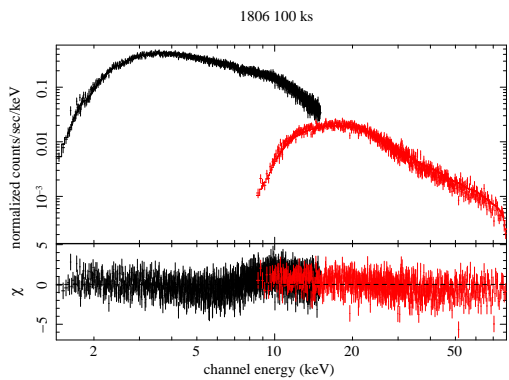}}
\caption{\footnotesize 
100 ks simulated Simbol X observation of \zerosei. The upper panel shows the model and the data while the lower panel shows the residuals with respect to the model. The data have been fitted with a single power-law.
}
\label{fig:1806_100k}
\end{figure}

As can be seen from Fig. \ref{fig:1806_100k}, where a fit using a single absorbed power law is shown, 100 ks at least are necessary to detect the spectral break (the $\chi_{r}^{2}$ increases from 1.21 to 1.03 if a break is included).
The black body component, on the other hand, is not statistically needed. This is not surprising since the first detection
of this component has been obtained with a 50 ks long {\it XMM}/pn exposure, and the latter is still 4-5 times more
sensitive than Simbol-X at 2 keV, where the black body component is most prominent.

\begin{figure}[ht!]
\resizebox{\hsize}{!}{\includegraphics[angle=-90,width=7cm]{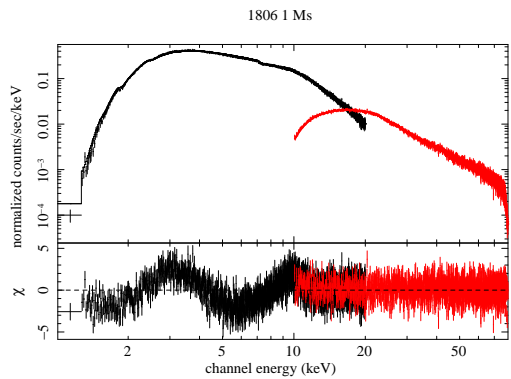}}
\caption{\footnotesize 
1Ms simulated Simbol X observation of \zerosei. The upper panel shows the model and the data while the lower panel shows the residuals with respect to the model. The data have been fitted with a single power law.
}
\label{fig:1806_1M}
\end{figure}

Increasing the exposure time to 1 Ms, see Fig. \ref{fig:1806_1M}, both the spectral break and the black body
component are evident if one looks at the residuals of the fit with a single absorbed power law. 
In fact the single power law fit is statistically not acceptable, yielding a $\chi_{r}^{2}$ of 3.3, which than decreases
to 1.6 including a break and finally to 1.02 if one includes the black body.

This shows us that with reasonable exposure times Simbol-X will provide us for the first time with a good quality spectrum for the SGRs as well as for AXPs. 

\section{Conclusions}
One of the major issues today in the magnetar's field is that no clear physical model has been developed yet, in order to explain the broad band spectrum of AXPs and SGRs. An important reason for this is that with the current instrumentation
it is not possible to obtain good quality broad band (0.1--200 keV) spectra. For every astrophysical object the broad
band coverage is nowadays an important issue, but the unexpected discovery of very hard tails in AXPs is a unique feature, and deserves particular attention in this sense.

We have shown that with Simbol-X we will be able to monitor for the first time all the spectral components of
these variable objects at once. In particular we will be able to pinpoint the differences between AXPs and SGRs, that
before the IBIS/ISGRI hard X-ray observations looked very similar for what concerns their persistent emission. We
will also be able to understand what is the driver of spectral changes in magnetars, i.e. if the changes below 10 keV
are in reality caused by the variation of an underlying high-energy component.

Concerning the timing, pulsations in the hard X-ray band have been detected for a few AXPs and possibly for
\zerosei~ (G\"otz et al., 2007, in preparation). Their detection needs long integration times, and the combination
of different instruments on different satellites, which introduces unavoidable uncertainties in the estimation of the pulsed
fraction. Simbol -X, being a focusing telescope with low background at hard X-rays, will surely reduce by an order
of magnitude our uncertainties and the exposure times required. In addition it will allow us to do broad band
phase resolved spectroscopy.

The detection of resonant cyclotron absorption lines would be the direct confirmation of the presence of  a huge magnetic
field in magnetars. Up to now, despite the deep searches below 10 keV, no firm detection of these features has been reported. One interesting feature around 6--15 keV has been reported by \citet{iwasawa} using {\it Ginga} data, during
the outburst of  the AXP 1E 2259+586. Being this feature close to the upper energy boundary of the instrument, its detection
is questionable. Simbol-X could clearly settle an issue like that.

A few of the AXPs are transients: this is a new subclass of magnetars that undergo large outbursts lasting years where
their flux changes by more than 3 orders of magnitude \citep[e.g.][]{cxo}. The mechanisms of these long outbursts is not known, and in addition due to several constraints, these sources could not be observed efficiently during their outbursts with
hard X-ray telescopes. Such an outburst (but also the quiescent phase) 
observed at hard X-rays with Simbol-X could surely help to shed some light
of the mechanisms that drive these long lived outbursts.
 
\begin{acknowledgements}
D.G. acknowledges the French Space Agency (CNES) for financial support. ISGRI has been realized and maintained in flight by CEA-Saclay/DAPNIA with the support of CNES.
\end{acknowledgements}

\bibliographystyle{aa}

\end{document}